\documentclass{elsart3}

\usepackage{natbib}

\usepackage{graphics}
\usepackage{graphicx}
\usepackage{epsfig}

\usepackage{amssymb}

\renewcommand{\deg}{^{\circ}}

\begin{document}

\begin{frontmatter}



\title{A dichotomy in radio jet orientations\thanksref{n:title}}

\thanks[n:title]{Based on observations with the NASA/ESA Hubble Space Telescope obtained at the Space Telescope Science Institute, which is operated by the Association of Universities for Research in Astronomy, Inc., under NASA contract NAS5-26555}

\author{Gijs Verdoes Kleijn}

\address{ESO, Garching bei M\"unchen, Germany (current address: Kapteyn Astronomical Institute, Postbus 800, 9700 AV, Groningen, The Netherlands)}

\author{Tim de Zeeuw}

\address{Leiden Observatory, Postbus 9513, 2300 RA, Leiden, The Netherlands}

\begin{abstract}
We examine the relative orientations of radio jets, central dust and stars in low-power (i.e., FR I and FR I/II) radio galaxies. We use the position angles of jet and dust to constrain the three-dimensional angle $\theta_{\rm DJ}$ between jet and dust. 
For galaxies with filamentary dust 'lanes' (which tend to be misaligned with the galaxy major axis) the jet is approximately perpendicular to the
dust structure, while for galaxies with elliptical dust distributions (typically aligned with the galaxy major axis) there is a much wider
distribution of $\theta_{\rm DJ}$.
The dust ellipses are consistent with being nearly
circular thin disks viewed at random viewing angles. The
lanes are likely warped, unsettled dust structures.
We consider two scenarios to explain the dust/jet orientation dichotomy. 
\end{abstract}

\begin{keyword}


\end{keyword}

\end{frontmatter}

\section{Introductory remarks}
\label{s:intro}
Since the late 1970s, many studies have found the position angles (PA) of jets in dusty radio
galaxies to be roughly perpendicular to the PA of the longest axis of the dust
structures (e.g., Kotanyi \& Ekers 1979; M\"{o}llenhoff, Hummel
\& Bender 1992; van
Dokkum \& Franx 1995; de Koff \etal\ 2000; de Ruiter \etal\ 2002). Capetti \& Celotti (1999) and Sparks
\etal\ (2000) reported that also the intrinsic,
i.e., three-dimensional, orientation of radio jets is roughly
perpendicular to the dust in samples of radio
galaxies with regular dust disks. In contrast, Schmitt \etal\ (2002) found in 
a sample of 20 radio galaxies with regular dust disks that the jets
are {\sl not} roughly perpendicular to the disks in three-dimensional
space.

Here we summarzie our analysis of the intrinsic orientation of dust and jets in 47 FR-I and FR-I/II radio galaxies using HST/WFPC2 broad-band imaging to explore the cause of the conflicting results on the dust-jet orientation. A detailed analysis is presented in Verdoes Kleijn \& de Zeeuw (2005). This paper also compares the dust properties of radio galaxies and quiescent ellipticals.

\section{Dust and jets: projected properties}
\label{s:dustproperties}
\begin{figure}
\resizebox{1.1\hsize}{!}{\includegraphics{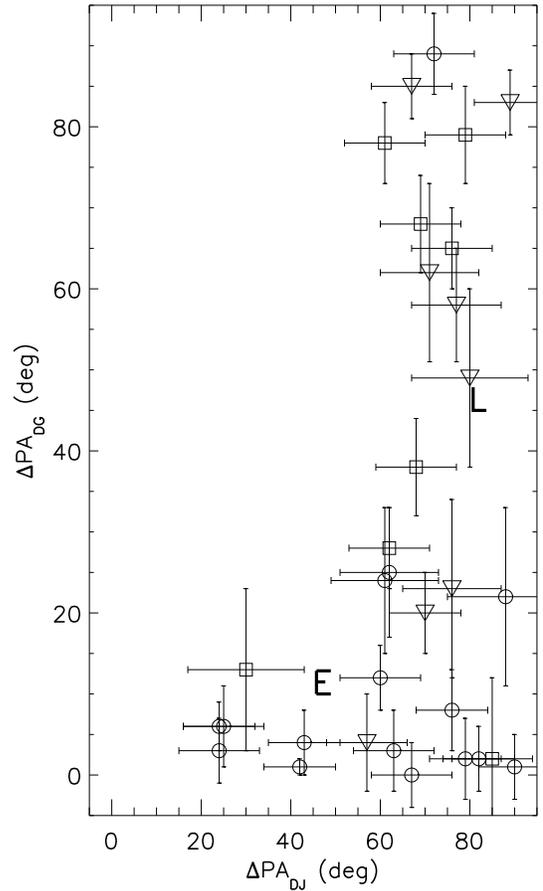}}
\caption{
PA difference $\Delta$PA$_{\rm DG}$
between dust axis and host galaxy major axis as a function of position angle difference
  between dust and radio jet $\Delta$PA$_{\rm DJ}$. Galaxies with a dust ellipse, a dust lane
  or an 'intermediate' morphology, i.e., in between lane and ellipse, are
  denoted by circles, triangles and squares, respectively. The regions 'E' and 'L' are populated mainly by dust ellipses and lanes respectively.
}
\label{f:dpa}
\end{figure}

The PA of the dust axis (PA$_{\rm D}$) is defined as the PA of the largest linear extent of the
dust feature. The PA of the stellar
isophotal major axis (PA$_{\rm G}$) is measured just outside the radius of the main dust distribution. The jet axis defines the PA$_{\rm J}$ for the radio component. Figure~\ref{f:dpa} shows three special features in the relative orientation of the three axes. First, no radio jets are observed close to the
dust axis (i.e., $\Delta$PA$_{\rm DJ}<20\deg$). Second, the
data points are distributed roughly along two regions forming a mirrored 'L' shape . Dust
structures in region E are roughly aligned with the galaxy major axis
($\Delta$PA$_{\rm DG} < 20\deg$) and have a wide distribution in relative
angles with the radio jet ($\Delta$PA$_{\rm DJ}\sim [20\deg-90\deg]$).
In contrast, dust features in region L are misaligned from the galaxy major
axis ($\Delta$PA$_{\rm DG} > 20\deg$) and have a narrow distribution in PA
differences with the radio jets ($\Delta$PA$_{\rm DJ}\sim
[60\deg-90\deg]$). These dust structures, while
misaligned with the galaxy, are in a very rough sense perpendicular to
the radio jets. Third, dust morphology correlates with these distinct classes of relative orientations. The appearance of dust structures in the radio galaxy sample varies from regular elliptical shapes  -- evoking the idea of
inclined disks -- via filamentary dust 'lanes' to highly irregular structures --suggesting unsettled
dust (Verdoes Kleijn \etal\ 1999; 2005). The dust ellipses tend to fall in region E and the dust lanes in region L.

\section{Intrinsic dust-jet orientations}
The relative intrinsic orientation between jet and dust is
characterized by the 'misalignment angle' $\theta_{\rm DJ}$ which is the angle between the jet and dust disk rotation axis. We have analysed the distribution of $\theta_{\rm DJ}$ separately for dust ellipses and lanes. The reason is that, besides the difference in $\Delta$PA$_{\rm DG}$, there are various other pieces of evidence that filamentary dust lanes are not simply the edge-on counterparts of dust ellipses (Verdoes Kleijn \etal\ 2005).

\begin{figure}
\resizebox{1.0\hsize}{!}{\includegraphics{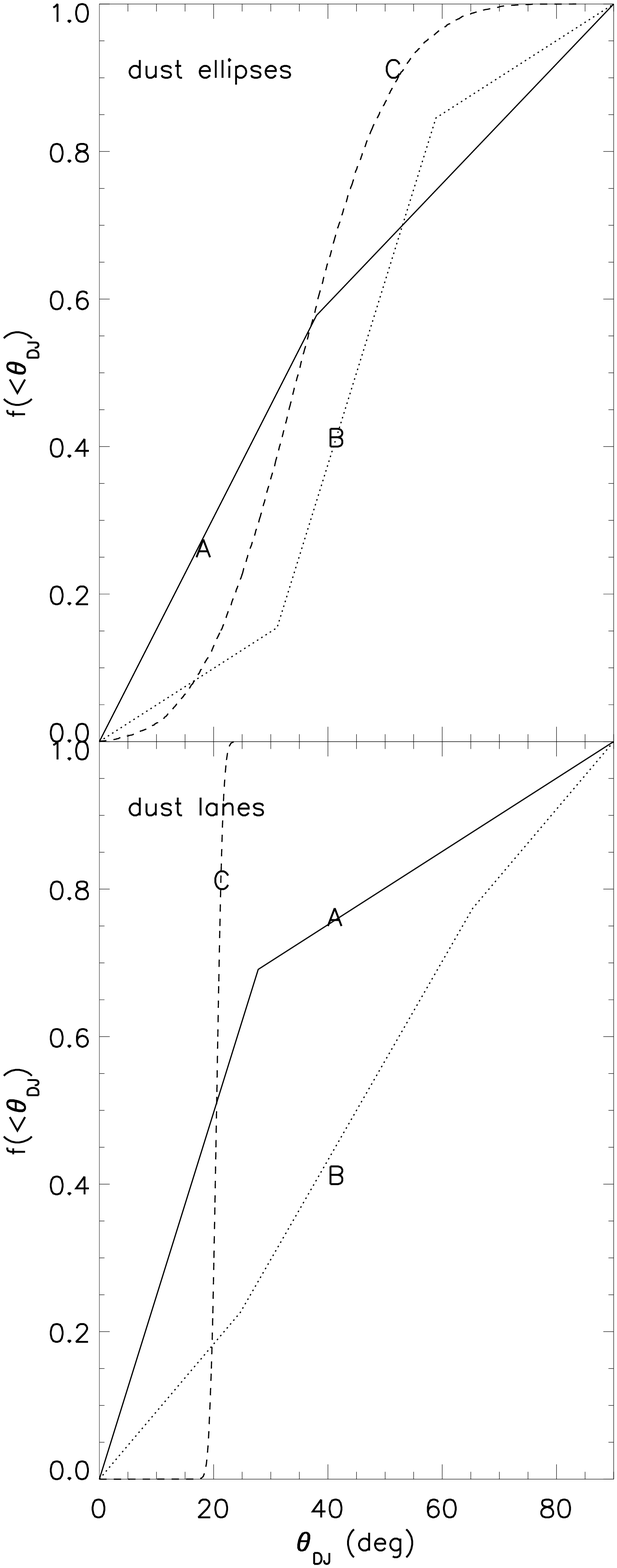}}
\caption{Cumulative distributions of the jet-dust misalignment angle $\theta_{\rm DJ}$ for
  the radio galaxies with dust ellipses (top) and dust lanes (bottom) for model A (solid curve), B (dotted) and C (dashed). The optimal parameters for the models are determined from a best fit to the observed (i.e., projected) relative orientations of dust and jets. Randomly oriented circular dust disks are assumed for the dust ellipses, and edge-on disks for lanes. Jets of radio galaxies with dust lanes have on average smaller misalignment angles than those in radio galaxies with dust ellipses.}
\label{f:psi_m}
\end{figure}

For dust ellipses, four of the 16 galaxies with measured $\Delta$PA$_{\rm DJ}$ require minimal $\theta^{\rm min}_{\rm
DJ} > 40\deg$. In other words, significant misalignments occur.  
Given the small number of observations, we constrain the distribution of misalignment angles $P_{\rm DJ}(\theta_{\rm DJ})$ further by exploring three parameterizations for $P_{\rm DJ}$ instead of
attempting a parameter-free recovery. In model A, $P_{\rm DJ}$ is assumed to be a single step-function to test the hypothesis that the distribution of $\theta_{\rm DJ}$ peaks at small or large angles. Model B assumes it to be a two-step function to test for peaks at intermediate misalignment angles (i.e., $\theta_{\rm DJ} = 45\deg$). Lastly we explore model C which assumes a Gaussian distribution truncated to the physically allowed region $0\deg \leq \theta_{\rm DJ} \leq 90\deg$ to explore arbitrary peaks and widths.  These models are fitted to the observed relative orientations of dust and jets using Maximum Likelihood optimization and results are shown in Figure~\ref{f:psi_m}.

The jet misalignment angle distribution in galaxies with dust ellipses is consistent with having a peak around $\theta_{\rm DJ} \sim 45\deg$. The width of this peak is not well constrained. The limited data set cannot rule out a spherically random distribution of misalignment angles or a very narrow peak at more than $95\%$ confidence. These conclusions do
not depend critically on either the assumed parameterization for $P_{\rm DJ}$ or the assumed thickness or ellipticity of the disk. Regardless of the assumed model, typically at least half of the radio jets make an angle of $45\deg$ or more with the symmetry axis of the dust disks. 

To constrain the three models for dust lanes, we assume that the lanes are exactly edge-on
systems to obtain an upper-limit on the typical $\theta_{\rm DJ}$. Figure~\ref{f:psi_m} shows that the upper limit to the median misalignment
angle for dust lane radio galaxies is smaller than the median angle for dust disk radio galaxies. 
The fact that dust lanes seem to be viewed close to edge-on prompts the
question: where are the relatively face-on dust lanes? We conclude from further analysis that the dust classified as either irregular or intermediate between ellipse and lane most likely represent the close to face-on counterparts of the edge-on lanes (Verdoes Kleijn \& de Zeeuw 2005). 

There are two scenarios to explain the dust-galaxy-jet orientation dichotomy for radio galaxies. 
Scenario I: the radio jets exert a torque on the nuclear dust in active galaxies which in some cases overcomes the gravitational torque forcing the nuclear dust in a plane perpendicular to the jets.
This scenario does not explain why the dust/galaxy misalignment is also seen in galaxies without radio jets. Scenario II: the angular momentum vector of unsettled nuclear dust is initially aligned with the radio jet in active galaxies but this alignment is lost as dust settles in an equilibrium plane of the galaxy. A weakness of this scenario is that the settling time-scales of the dust seem short compared to the age of the radio sources. In conclusion, the analysis establishes a dichotomy in jet orientations related to distinct dust morphologies, but the current data set is too small to distinguish between scenarios explaining the phenomenon.










\end{document}